\documentclass[aps,prb,superscriptaddress,reprint]{revtex4-2}
\newcommand{\VERSION}{resubmit}  %% FROZEN -  RESUBMIT SEE DATE COMMAND
    
\usepackage[utf8]{inputenc}
\usepackage{graphicx,dcolumn,bm,amsmath,amssymb,euscript}

\graphicspath{{figures/vaccept/}}

\usepackage{color, colortbl}   %% in text, use as \textcolor{green}{text that shows up green}
\usepackage[bookmarksopen,bookmarksnumbered,colorlinks,citecolor=red,urlcolor=blue,filecolor=green]{hyperref}

\begin{document}

\title{Crystal Truncation Rods from Miscut Surfaces with Alternating Terminations}

\author{Guangxu Ju}
    \email[correspondence to: ]{juguangxu@gmail.com}
	\altaffiliation[current address: ]{Lumileds Lighting Co., San Jose, CA 95131 USA.}
	\affiliation{Materials Science Division, Argonne National Laboratory, Lemont, IL 60439 USA}
\author{Dongwei Xu}
	\affiliation{Materials Science Division, Argonne National Laboratory, Lemont, IL 60439 USA}
	\affiliation{School of Energy and Power Engineering, Huazhong University of Science and Technology, Wuhan, Hubei 430074, China}
\author{Carol Thompson}
	\affiliation{Department of Physics, Northern Illinois University, DeKalb, IL 60115 USA}
\author{Matthew J. Highland}
	\affiliation{X-ray Science Division, Argonne National Laboratory, Lemont, IL 60439 USA}
\author{Jeffrey A. Eastman}
	\affiliation{Materials Science Division, Argonne National Laboratory, Lemont, IL 60439 USA}
\author{Weronika Walkosz}
	\affiliation{Department of Physics, Lake Forest College, Lake Forest, IL 60045 USA}
\author{Peter Zapol}
	\affiliation{Materials Science Division, Argonne National Laboratory, Lemont, IL 60439 USA}
\author{G. Brian Stephenson}
    \email[correspondence to: ]{stephenson@anl.gov}
	\affiliation{Materials Science Division, Argonne National Laboratory, Lemont, IL 60439 USA}

\date{revision \VERSION  : February 18, 2021}  % DATE OF SUBMIT FROZEN

\begin{abstract}
Miscut surfaces of layered crystals can exhibit a stair-like sequence of terraces having periodic variation in their atomic structure. For hexagonal close-packed and related crystal structures with an $\alpha \beta \alpha \beta$ stacking sequence, there have been long-standing questions regarding how the differences in adatom attachment kinetics at the steps separating the terraces affect the fractional coverage of $\alpha$ vs. $\beta$ termination during crystal growth. To demonstrate how surface X-ray scattering can help address these questions, we develop a general theory for the intensity distributions along crystal truncation rods (CTRs) for miscut surfaces with a combination of two terminations. We consider half-unit-cell-height steps, and variation of the coverages of the terraces above each step. Example calculations are presented for the GaN $(0 0 0 1)$ surface with various reconstructions. These show which CTR positions are most sensitive to the fractional coverage of the two terminations. We compare the CTR profiles for exactly oriented surfaces to those for vicinal surfaces having a small miscut angle, and investigate the circumstances under which the CTR profile for an exactly oriented surface is equal to the sum of the intensities of the corresponding family of CTRs for a miscut surface.
\end{abstract}

\maketitle
\section{Introduction}

Crystal truncation rods (CTRs) are features in the scattering patterns of crystals that arise from the truncation of the bulk crystal lattice at a surface \cite{1986_Robinson_PRB33_3830}.
They consist of streaks of scattering intensity extending away from all Bragg peaks in the direction normal to the crystal surface.
The intensity distributions along the CTRs provide a sensitive measure of the atomic structure of the surface.
In particular, they provide information about the
termination plane of the crystal and the reconstruction of surface layers \cite{2004_Fenter_JApplCryst37_977},
as well as the arrangement of atomic-scale steps on the surface \cite{2017_Petach_PRB95_184104}.
\textit{In situ} X-ray measurements of CTRs have been used to determine the atomistic mechanisms of crystal growth \cite{1988_Vlieg_PRL61_2241,1992_Vlieg_PRL68_3335,1992_Fuoss_PRL69_2791,1999_Stephenson_APL74_3326,2002_RamanaMurty_APL80_1809,2009_Ferguson_PRL103_256103,2014_Perret_APL105_051602}, such as the classical homoepitaxial growth mechanisms of 1-dimensional step flow, 2-dimensional island nucleation and coalescence, and 3-dimensional roughening \cite{1993_Tsao_MatFundMBE,2001_Giesen_ProgSurfSci68_1}.
Because of their penetrating nature and atomic-scale sensitivity, X-ray methods are uniquely suitable for \textit{in situ} studies in non-vacuum environments.
Here we calculate the behavior of CTR intensities to address a long-standing issue in crystal growth, the variation in the morphology of surfaces having alternating terminations and step types, driven by differences in step kinetics during step-flow growth. 

\begin{figure}
\includegraphics[width=1\linewidth]{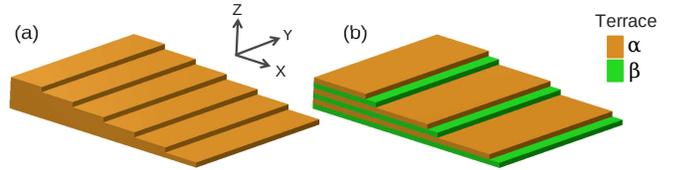}
\caption{Miscut surfaces with (a) identical terraces and (b) terraces with alternating structures. Alternating structures often arise for crystal structures with a screw axis normal to the surface. In this case half-unit-cell-height steps separate terraces with differing orientations of the terminating layer. The alternating structures and kinetic properties of the steps can lead to unequal terrace fractions.  \label{fig:termination}}
\end{figure}

The simplest atomic model of a miscut (vicinal) crystal surface consists of a stair-like sequence of identical terraces, separated by identical steps, typically of full-unit-cell height, as shown in Fig.~\ref{fig:termination}(a). 
The average step spacing is determined by the miscut angle of the surface (i.e. the angle of the surface away from the closest low-index crystallographic orientation).
During stable step flow growth, the terrace widths are typically all nearly equal, since all steps have identical kinetic properties. 
However, miscut surfaces can acquire more complex morphologies, with terraces separated by fractional-unit-cell-height steps, and step structures and properties that vary from step to step, as shown in Fig.~\ref{fig:termination}(b).
A typical case is a unit cell with a multiple-layer structure \cite{2004_Fenter_JApplCryst37_977}.
In particular, when the space group of the crystal includes a screw axis or a glide plane, the surface nearly perpendicular to this symmetry element can have a sequence of terraces with the same atomic arrangement, but different in-plane orientations.
This gives differences in the structure and kinetics of neighboring steps, leading to fundamentally new crystal growth characteristics that produce fascinating
terrace morphologies \cite{2004_vanEnckevort_ActaCrystA60_532}.

\begin{figure}
\includegraphics[width=1\linewidth]{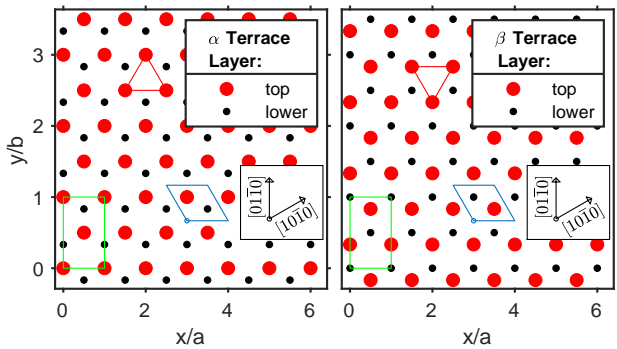}
\caption{Structure of $\alpha$ and $\beta$ terraces of the (0001) surface of an HCP-type crystal, e.g. the Ga sites in wurtzite-structure GaN. Red triangle of top-layer sites around $6_3$ screw axis shows difference between alternating $\alpha$ and $\beta$ layers. Blue rhombus shows conventional HCP unit cell; green rectangle shows orthohexagonal unit cell. Axes give coordinates in terms of orthohexagonal lattice parameters $a$ and $b = \sqrt{3} a$. \label{fig:alpha_beta}}
\end{figure}

\begin{figure}
\includegraphics[width=0.8\linewidth]{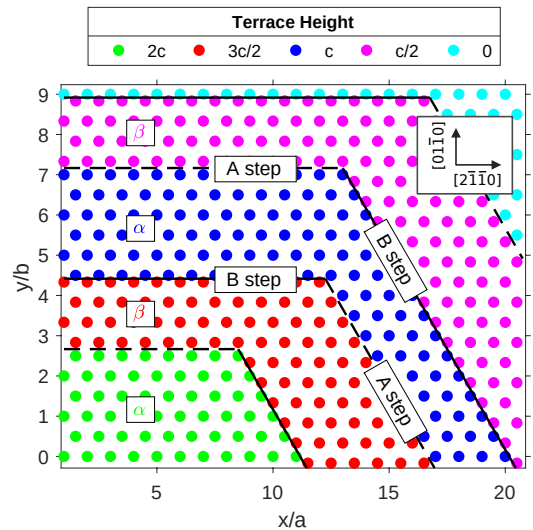}
\caption{Terrace and step structure of an $(0 0 0 1)$ surface of an HCP-type crystal with regions miscut in two different azimuths. Circles show in-plane positions of top-layer atoms on each terrace, with color indicating height. Orthohexagonal lattice parameters are $a$, $b$, and $c$. Steps of height $c/2$ typically have lowest edge energy when they are normal to $[0 1 \overline{1} 0]$, $[1 0 \overline{1} 0]$, or $[1 \overline{1} 0 0]$. Steps of a given azimuth have alternating structures, $A$ and $B$. The step structure changes from $A$ to $B$ or $B$ to $A$ when they change azimuth by $60^{\circ}$. 
\label{fig:A_B}}
\end{figure}

A commonly encountered version of this effect occurs on the basal-plane $\{0 0 0 1\}$-type surfaces of crystals having hexagonal close-packed (HCP) or related structures, which are normal to a $6_3$ screw axis.
The close-packed layers in HCP crystals have 3-fold symmetry alternating between $180^\circ$-rotated orientations from layer to layer, as shown by the $\alpha$ and $\beta$ terrace structures in Fig.~\ref{fig:alpha_beta}.
The $\alpha \beta \alpha \beta$ stacking sequence typically results in half-unit-cell-height steps on miscut surfaces.
Often the lowest energy steps are normal to $[0 1 \overline{1} 0]$-type directions.
The alternating structures of these steps are conventionally labelled $A$ and $B$ \cite{1999_Xie_PRL82_2749,2001_Giesen_ProgSurfSci68_1}
as shown on Fig.~\ref{fig:A_B}.
When the in-plane azimuth of an $A$ step changes by $60^\circ$, e.g. from $[0 1 \overline{1} 0]$ to $[1 0 \overline{1} 0]$, its structure changes to $B$, and \textit{vice versa}.
Differences in the dynamics of adatom attachment at $A$ and $B$ steps have strong effects on the surface morphology produced during growth.

Images of $\{0 0 0 1\}$ surfaces suggesting the alternating nature of the steps have been obtained for several HCP-related systems, including SiC \cite{1951_Verma_PhilMag42_1005,1979_Sunagawa_JCrystGrowth46_451,1982_vanderHoek_JCrystGrowth58_545}, GaN \cite{1999_Xie_PRL82_2749,1999_Heying_JAP85_6470,2001_Vezian_MatSciEngB82_56,2002_Zauner_JCrystGrowth240_14,2006_Xie_PRB_085314,2007_Krukowski_CrystResTechnol42_1281,2008_Zheng_PRB77_045303,2013_Turski_JCrystGrowth367_115,2013_Lin_ApplPhysExp6_035503}, AlN \cite{2017_Pristovsek_physstatsolb254_1600711}, and ZnO \cite{2002_Chen_APL80_1358}. 
Such images typically indicate a tendency for local pairing of steps (i.e. alternating step spacings), consistent with predictions that $A$ and $B$ steps can have significantly different energies and/or attachment kinetics \cite{1999_Xie_PRL82_2749,2006_Xie_PRB_085314,2013_Turski_JCrystGrowth367_115,2011_Zaluska-Kotur_JAP109_023515,2010_Zaluska-Kotur_JNoncrystSolids356_1935,2017_Xu_JChemPhys146_144702,2017_Chugh_ApplSurfSci422_1120,2020_Akiyama_JCrystGrowth532_125410,2020_Akiyama_JJAP59_SGGK03}.
In particular, different attachment kinetics at $A$ and $B$ steps can produce a tendency to step pairing during growth and thus to different local fractions of $\alpha$ and $\beta$ terraces.
In limiting cases, the $\alpha$ terrace fraction $f_\alpha$ can approach zero or unity, when pairs of half-unit-cell-height steps join to form full-unit-cell-height steps.
However, for $\{0 0 0 1\}$ surfaces of HCP-related systems it has been difficult to distinguish experimentally the terrace orientation, and thus to determine whether a given set of steps is of $A$ or $B$ type.
Thus it has not been possible to experimentally determine whether $A$ or $B$ steps have faster adatom attachment kinetics.
This can be solved by a method to distinguish the fraction of $\alpha$ and $\beta$ terraces, especially an \textit{in situ} measurement in the relevant growth environment.

Streaks of scattering intensity extending away from Bragg peaks normal to the surfaces of finite-sized crystals are inherent in exact treatments of X-ray scattering, extending back to early work \cite{james1954optical,1964_Batterman_RevModPhys36_681}. As synchrotron sources
enabled detailed study of these surface-sensitive features, theoretical treatments specific to CTRs from truncated perfect crystals were developed \cite{1985_Andrews_JPhysC18_6427,1986_Robinson_PRB33_3830}. Subsequent work included the effect of overlayers or reconstructions \cite{1988_Robinson_PRB38_3632,2004_Fenter_JApplCryst37_977}
and miscut surfaces \cite{1995_Held_PRB51_7262,1999_Munkholm_JApplCryst32_143,2002_Trainor_JApplCryst35_696,2007_Wollschlager_PRB75_245439,2017_Petach_PRB95_184104}. Previous general treatments of miscut surfaces typically considered only full unit cell steps; half-unit cell steps on Si $(0 0 1)$ have also been considered \cite{1999_Munkholm_JApplCryst32_143,2007_Wollschlager_PRB75_245439}.

Here we develop expressions for the CTR intensity distributions for surfaces with two terrace types, $\alpha$ and $\beta$.
We explicitly include the effects of surface reconstructions.
We consider straight steps, periodically spaced.
Surface roughness is accounted for by applying a roughness factor, as derived in prior work \cite{2017_Petach_PRB95_184104}.
New aspects of our calculations include a general treatment of half-unit-cell-height steps, with fractional surface coverage of two terrace types; generalization from a cubic to an orthorhombic bulk unit cell, which allows us to consider HCP type crystals using orthohexagonal coordinates;
and the effects of absorption and extinction.
We use a mathematical formalism that emphasizes how exact results can be obtained through the summation of geometric sequences.
We compare the CTR profiles for exactly-oriented and miscut surfaces, and consider the relationships between the two cases.

To provide a concrete example, we make calculations for the GaN $(0 0 0 1)$ surface, for which we have recently carried out a surface X-ray scattering study \cite{2020_Ju_NC}. 
We show how the CTRs vary as a function of the surface fraction of $\alpha$ vs $\beta$ terraces. 
This demonstrates that surface X-ray scattering can be used to distinguish the fraction of the surface covered by $\alpha$ or $\beta$ terraces during growth, and thus unambiguously determine differences in the attachment kinetics at $A$ and $B$ steps.

\section{Surface X-ray Scattering Theory}

In this section we develop expressions for the intensity distributions along CTRs and demonstrate their sensitivity to the surface termination arrangements.
The X-ray reflectivity along the CTRs can be calculated by adding the complex amplitudes from the bulk crystal and the terminating overlayers, with proper phase relationships.
We start with the simple case of an exactly-oriented surface, and then extend this to the case of a vicinal surface having an array of straight steps.
After developing general expressions, we show calculations for the GaN $(0 0 0 1)$ surface for both cases.

\begin{figure*}
\includegraphics[width=1\linewidth]{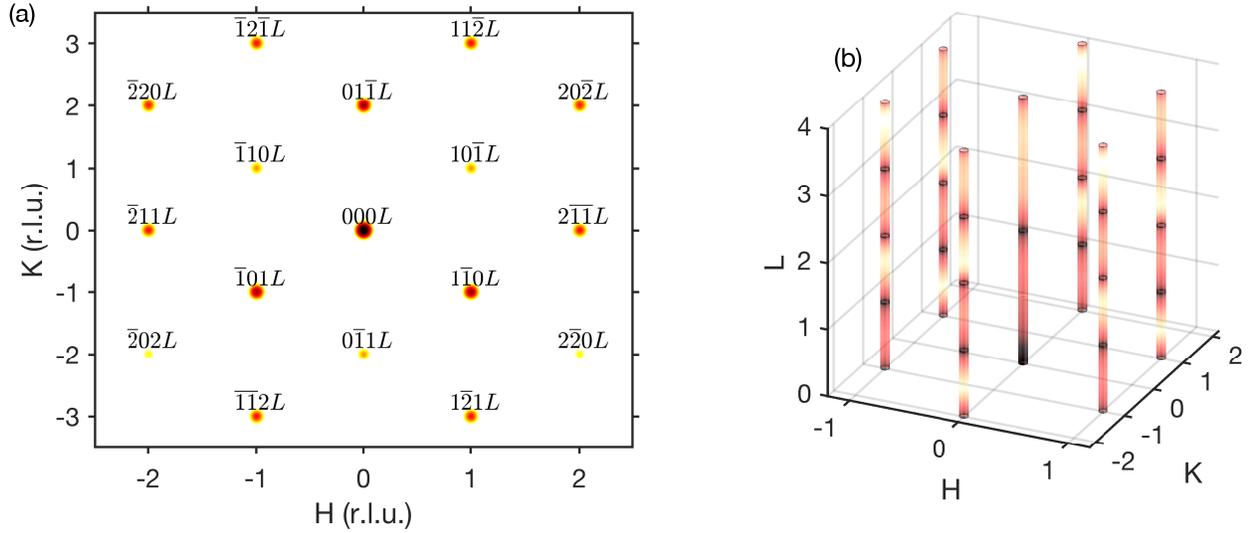}
\caption{(a) Hexagonal pattern of CTRs at fixed $L = 0.5$, for a GaN basal plane surface. Axes give $HK$ in orthohexagonal units; Miller-Bravais indices are listed by each CTR. (b) Reciprocal space map of CTRs for a GaN basal plane surface. Color indicates variation in reflectivity as a function of $L$. Axes give $HKL$ in orthohexagonal units.  \label{fig:HKL}}
\end{figure*}

\begin{figure*}
\includegraphics[width=1\linewidth]{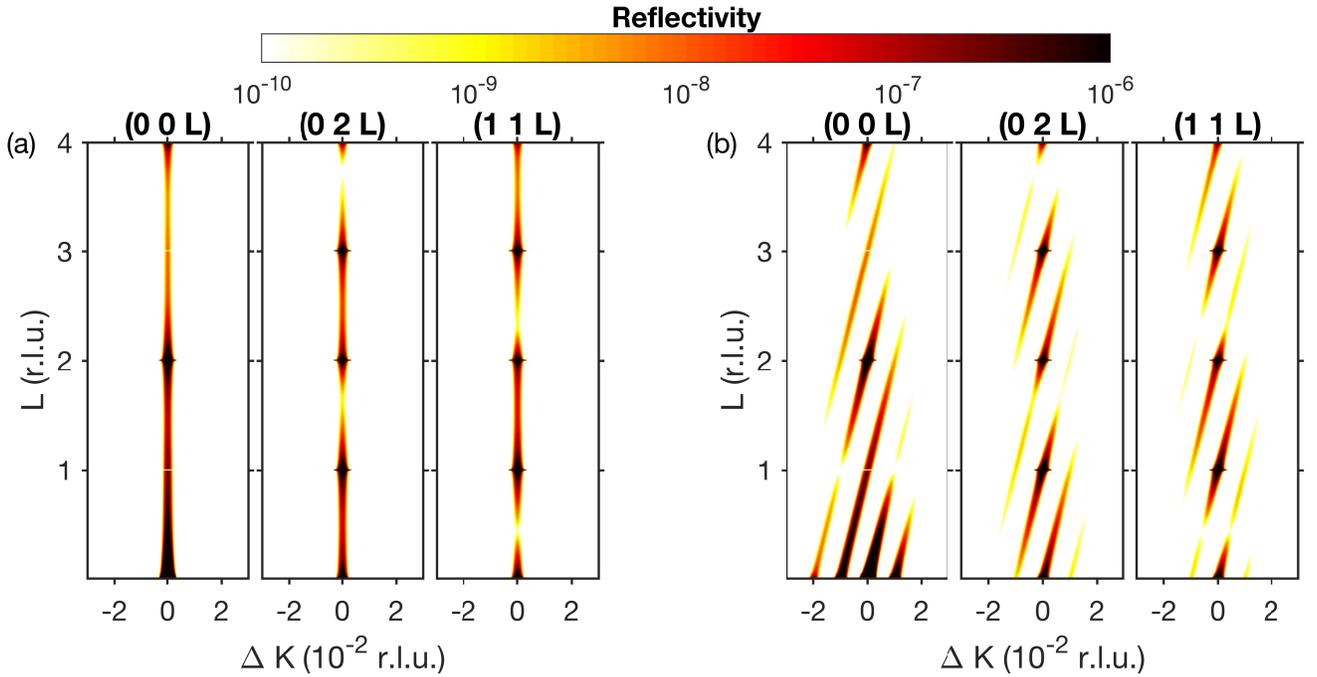}
\caption{Color indicates reflectivity of CTRs as a function of $L$ and $\Delta K \equiv K - K_0$. Width of CTRs in $K$ is arbitrarily set to $0.001$ r.l.u. (a) CTRs for an exactly oriented GaN $(0 0 1)$ surface with no reconstruction and $f_\alpha = 0$, for various $H_0$ and $K_0$ shown at top. (b) CTRs for a miscut GaN $(0 0 1)$ surface with $M = 100$, no reconstruction, and $f_\alpha = 0$, for various $H_0$, $K_0$, and $L_0$. Orthohexagonal indices are used for $H K L$.
\label{fig:KL}}
\end{figure*}

Although our motivation for this work is based on hexagonal crystals, it is more convenient to use the Cartesian coordinates of an orthorhombic unit cell.
As shown in Figs.~\ref{fig:alpha_beta} and \ref{fig:HKL}, the hexagonal lattice can be mapped to an orthorhombic lattice using a doubled unit cell and an orthohexagonal coordinate system \cite{1965_Otte_PhysStatSol9_441}. 
The lattice parameters of the orthorhombic unit cell are $a$, $b$, and $c$ in the $x$, $y$, and $z$ directions, respectively.
The corresponding Cartesian components of the scattering wavevector $\mathbf{Q}$ are given by $(Q_x,Q_y,Q_z)$,
which are related to reciprocal lattice units $HKL$ by
$Q_x = (2 \pi/a) H$, $Q_y = (2 \pi/b) K$, $Q_z = (2 \pi/c) L$.
Bragg peaks locations have integer indices $H_0 K_0 L_0$ in reciprocal lattice units. 

For a hexagonal crystal, the orthohexagonal coordinate system has a lattice parameter $b \equiv a \sqrt{3}$, where $a$ is the in-plane lattice parameter of the conventional hexagonal unit cell, as shown in Fig.~\ref{fig:alpha_beta}.
The out-of-plane lattice parameter $c$ is the same in both coordinate systems.
This gives Cartesian $x$, $y$, and $z$ axes parallel to the hexagonal $[2 \overline{1} \overline{1} 0]$, $[0 1 \overline{1} 0]$, and $[0 0 0 1]$ directions, respectively.
In reciprocal space, the orthohexagonal coordinates $H K L$ are related to the hexagonal Miller-Bravais reciprocal space coordinates $h k i \ell$ by $H = h$, $K = h+2k$, $L = \ell$, as shown in Fig.~\ref{fig:HKL}.
In particular, the hexagonal $(1 0 \overline{1} 0)$ and $(0 1 \overline{1} 0)$ correspond to the orthohexagonal $(1 1 0)$ and $(0 2 0)$, respectively.
Because of the doubled unit cell, orthohexagonal indices $H$ and $K$ must both be even or odd integers at allowed Bragg peaks.
For the remainder of this paper, we will use 3-index orthohexagonal indices $H K L$ rather than 4-index hexagonal Miller-Bravais indices $h k i \ell$.

\subsection{Exactly oriented surface with two terminations}

For an exactly oriented surface normal to the $z$ direction, the CTRs extend continuously in the $Q_z$ direction at fixed $Q_x$ and $Q_y$ through each Bragg peak.
The CTRs from the Bragg peaks of different $L_0$ at the same $H_0 K_0$ all overlap, as shown in Fig.~\ref{fig:KL}(a).

The contribution to the complex amplitude $r_\mathrm{bulk}$ of the reflectivity from the truncated bulk crystal below the terminating overlayers can be obtained by coherently summing the scattering from each unit cell in a column normal to the surface, taking into account a factor $Z$ difference between the scattering amplitude from each unit cell \cite{1988_Robinson_PRB38_3632,1997_Thompson_APL71_3516}.
The sum of this geometrical sequence is given by
\begin{equation}
    r_\mathrm{bulk} = r_\mathrm{f} \, F_\mathrm{bulk} \sum_{m = -\infty}^0 Z^m = r_\mathrm{f} \, F_\mathrm{bulk} \, \frac{Z}{Z - 1},
    \label{eq:r_b}
\end{equation}
where $r_\mathrm{f} \equiv 4 \pi i r_0 / A Q$, $r_0 = 2.817 \times 10^{-13}$~cm is the Thomson radius of the electron, $A$ is the in-plane area of the unit cell, and $Q$ is the magnitude of the wavevector.  
The bulk structure factor $F_\mathrm{bulk}$ is
\begin{equation}
    F_\mathrm{bulk} = \sum_k g_k(Q) \sum_n \exp(i \mathbf{Q} \cdot \mathbf{r}_{kn}^\mathrm{bulk}).
\end{equation}
Here the first sum is over the $k$ chemical elements present in the crystal, $g_k(Q) \equiv f_k(Q) \exp(-u_k^2 Q^2 / 2)$ is the atomic form factor $f_k(Q)$ modified by the Debye-Waller thermal vibration length $u_k$ for element $k$, the second sum is over the $n$ bulk atoms of type $k$ in a unit cell, and $\mathbf{r}_{kn}^\mathrm{bulk}$ is the position of bulk atom $n$ of type $k$.
Unlike Bragg peak intensities for bulk crystals, CTR profiles depend upon the choice of the boundaries of the unit cell.
Shifting some atoms between the top and bottom of the unit cell by a lattice translation does not change $F_\mathrm{bulk}$ at the $\mathbf{Q}$ of Bragg peaks, but it does change $F_\mathrm{bulk}$ at $\mathbf{Q}$ values along the CTR between the Bragg peaks.
The unit cell boundaries used for $\mathbf{r}_{kn}^\mathrm{bulk}$ determine the surface termination, which generally affects the CTR profile \cite{2004_Fenter_JApplCryst37_977}.

The quantity $Z$ in Eq.~(\ref{eq:r_b}) is the ratio of the scattering amplitude from one unit cell to that from the unit cell at $\Delta z = -c$ below it.
It consists of a phase factor and an absorption factor,
\begin{align}
    Z & \equiv \exp (i Q_z c) \, \exp (\epsilon c / Q_z) \nonumber \\
    &= \exp (2 \pi i L) \, \exp (\epsilon c^2 / 2 \pi L),
    \label{eq:Z}
\end{align}
where $\epsilon = 4 \pi / (\lambda \ell_\mathrm{abs})$ is related to the photon wavelength $\lambda$ and absorption length $\ell_\mathrm{abs}$.
Typically the absorption factor is very close to unity and only becomes important in $Z - 1$ when the phase factor is unity, e.g. at Bragg peaks.
One can see from Eq.~(\ref{eq:r_b}) that the reflectivity amplitude is built up by summing the scattering from each layer of the semi-infinite crystal in the $z$ direction from $m = -\infty$ to $m = 0$.

\begin{figure}
\includegraphics[width=1\linewidth]{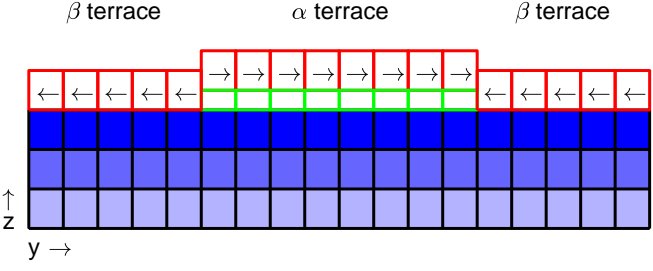}
\caption{Bulk unit cells (black) and reconstructed unit cells (red), for an exactly oriented surface with an island of $\alpha$ termination in a $\beta$ matrix.
The extra half unit cells producing the shift from the $\beta$ to the $\alpha$ termination are shown in green.
Blue shade indicates index $m$ of sum in Eq.~(\ref{eq:r_b}), with final term $m = 0$ darkest. \label{fig:exact}}
\end{figure}

To provide a direct comparison with the results for miscut surfaces obtained below, we allow the surface to have a mixture of two terminations, $\alpha$ and $\beta$. 
For an exactly oriented surface, this consists of islands of $\alpha$ in a matrix of $\beta$, or vice versa, as shown in Fig.~\ref{fig:exact}.
We also allow reconstructions in which the atoms in the top layer of unit cells at the surface are relaxed from their bulk crystal positions, and there can be extra atoms bonded to the surface.
The reconstructed reflectivity amplitude per unit area of the $x = \alpha$ or $\beta$ overlayer is
\begin{equation}
    r_\mathrm{rec}^x = r_\mathrm{f} \,\, F_\mathrm{rec}^x \,\, Z, 
\end{equation}
where the structure factor of the reconstruction $F_\mathrm{rec}^x$ is
\begin{equation}
    F_\mathrm{rec}^x = \sum_j \theta_{xj} \sum_k g_k(Q) \sum_n \exp(i \mathbf{Q} \cdot \mathbf{r}_{jkn}^x).
\end{equation}
Here the first sum is over the $j$ possible domain orientations of the reconstruction, $\theta_{xj}$ is the fraction of domain $j$ for the $x = \alpha$ or $\beta$ component, the second sum is over the $k$ chemical elements present in the reconstruction, the third sum is over the $n$ atoms of type $k$ in a unit cell, and $\mathbf{r}_{jkn}^x$ is the position of atom $n$ of type $k$ in domain orientation $j$ for the $x = \alpha$ or $\beta$ component.
The total reflectivity amplitude is the sum of the complex amplitudes from the bulk and both reconstructed components,
\begin{equation}
    r_\mathrm{tot} = r_\mathrm{bulk} + f_\alpha r_\mathrm{rec}^\alpha + (1 - f_\alpha) r_\mathrm{rec}^\beta,
    \label{eq:rt}
\end{equation}
where $f_\alpha$ is the fraction of the surface covered by the $\alpha$ termination.

\begin{figure}
\includegraphics[width=1\linewidth]{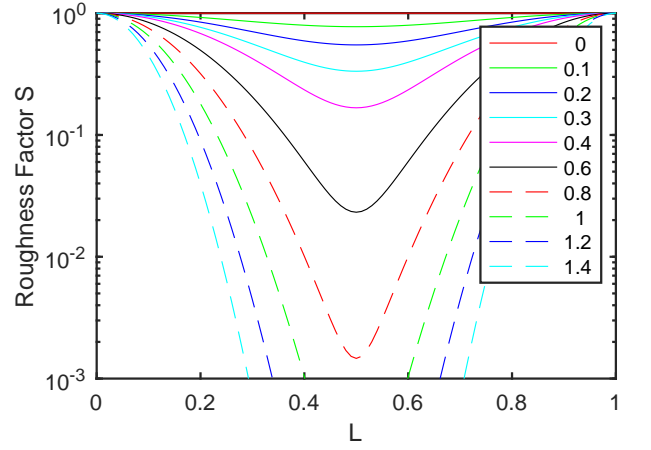}
\caption{Roughness factor $S$ as a function of $L$, for various values of $\sigma_R / c$ given in the legend. \label{fig:S}}
\end{figure}

The total reflectivity amplitudes calculated above are for the kinematic limit in which they are much smaller than unity.
Near the Bragg peaks, where the reflectivity amplitude approaches unity, the kinematic expressions can be corrected using \cite{1997_Thompson_APL71_3516}
\begin{equation}
    r_\mathrm{tot}^{dyn} = \frac{2 r_\mathrm{tot}}{1 + \sqrt{1 + 4 |r_\mathrm{tot}|^2}},
    \label{eq:dyn}
\end{equation}
which insures that the amplitude of the reflectivity does not exceed unity.
The intensity reflectivity is the square of the modulus of the amplitude reflectivity,
\begin{equation}
    R = |r_\mathrm{tot}^{dyn}|^2 \, S,
    \label{eq:R}
\end{equation}
where the final factor $S$ has been introduced to account for surface roughness.
Here we use a form for the roughness factor as a function of $L$ for the exactly oriented surface adopted from that obtained for a miscut surface \cite{2017_Petach_PRB95_184104},
\begin{align}
    S &= \sin^2(\pi L) \sum_{L_0 =  -\infty}^\infty \frac{\exp \big [-\sigma_R^2 (2\pi/c)^2 (L - L_0)^2 \big ]}{\pi^2 (L - L_0)^2} \nonumber \\
    & = 1 - \sin^2(\pi L) \left ( \frac{4 \sigma_R}{\pi^{1/2} c} - \sum_{n = 1}^\infty \cos(2 \pi n L) \times \right . \nonumber \\
    &\left . \left [ 4 n \, \mathrm{erfc} \left ( \frac{nc}{2\sigma_R} \right ) - \frac{8\sigma_R}{\pi^{1/2} c} \exp \left ( - \frac{n^2c^2}{4\sigma_R^2} \right ) \right ] \right ) \nonumber \\
    &\approx 1 - (4 \sigma_R /\pi^{1/2} c)  \sin^2(\pi L),
    \label{eq:S}
\end{align}
where $\sigma_R$ is the surface roughness (related to the total step roughness given in \cite{2017_Petach_PRB95_184104} by $\sigma_R = c \, \tilde{\sigma}_{tot}$), and the final approximation holds for $\sigma_R < 0.3 c$.
Figure~\ref{fig:S} shows plots of this function for various values of $\sigma_R/c$.

Away from the Bragg peaks, so that the dynamical scattering correction can be neglected, the intensity reflectivity can be written as
\begin{equation}
    R = |r_\mathrm{f}|^2 \, |Z|^2 \, \left | \frac{F_\mathrm{bulk}}{Z - 1} + f_\alpha F_\mathrm{rec}^\alpha + (1 - f_\alpha) F_\mathrm{rec}^\beta \right |^2 S.
    \label{eq:R2}
\end{equation}

\begin{figure*}
\includegraphics[width=0.8\linewidth]{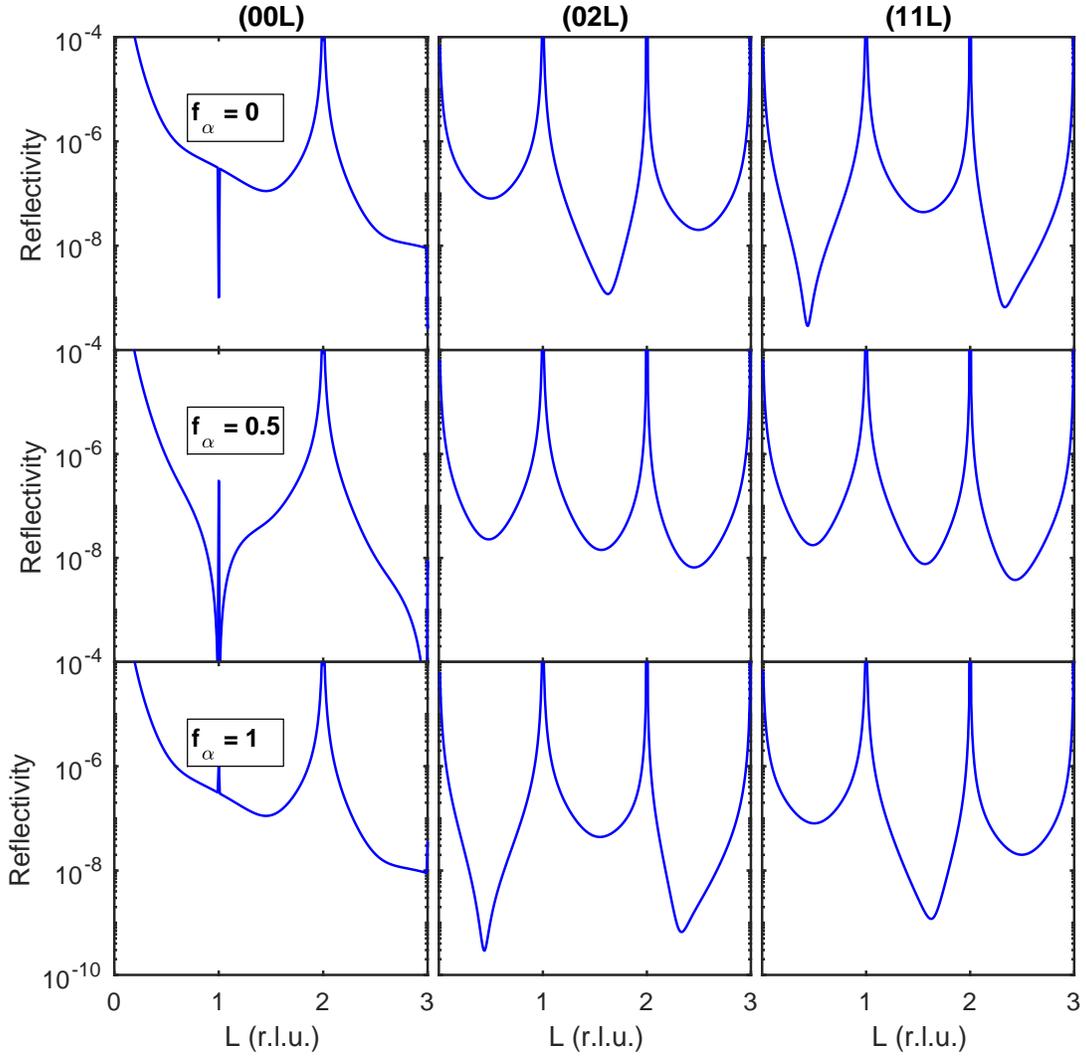}
\caption{Calculated reflectivity as a function of $L$ of CTRs for an exactly oriented GaN $(0 0 1)$ surface with no reconstruction. Left column: (0 0 L); middle column: $(0 2 L)$; right column: $(1 1 L)$. Values of $f_\alpha$ are given for each row. \label{fig:CTR_exact}}
\end{figure*}

\begin{figure*}
\includegraphics[width=1.0\linewidth]{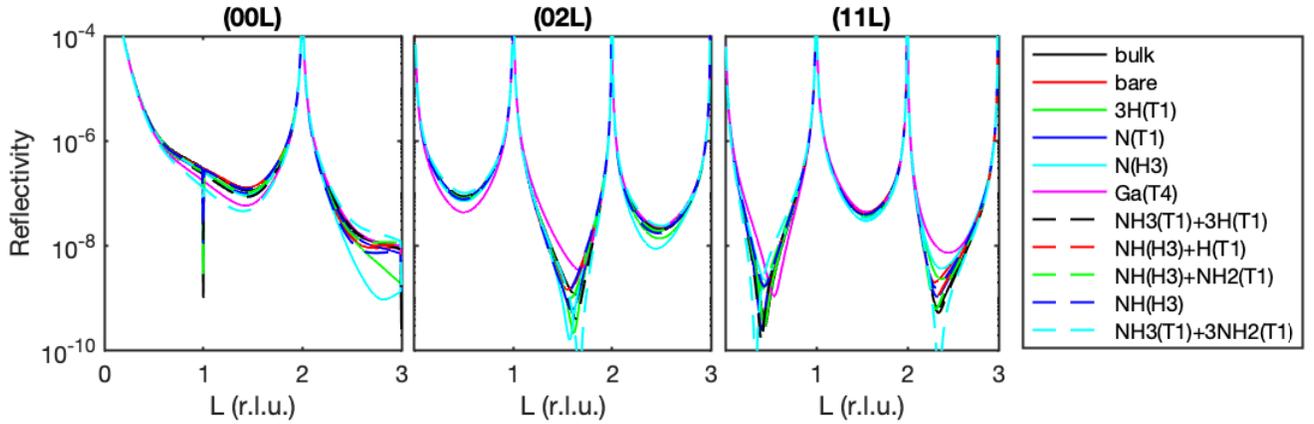}
\caption{Calculated reflectivity of CTRs for an exactly oriented GaN $(0 0 1)$ surface with various reconstructions, for $f_\alpha = 0$. \label{fig:CTR_exact_recon}}
\end{figure*}

\subsection{Exactly oriented GaN $(0 0 1)$ surface}

To illustrate the behavior for the basal plane of an HCP-type system, we consider Ga-face GaN $(0 0 1)$ surfaces with a Ga termination for the bulk, so that the top-layer atoms shown in Fig.~\ref{fig:A_B} are Ga.
Table~\ref{tab:coord_sub} in Appendix A lists the atomic coordinates used for the bulk.
We use atomic coordinates for the reconstructed overlayers calculated previously \cite{2012_Walkosz_PRB_85_033308}.
Since these were calculated using a $2 \times 2$ unit cell, for consistency the unit cell sums used in calculating the structure factors here are carried out over two adjacent orthohexagonal unit cells having an area $A = 2 a b$, which is normalized out in the denominator of $r_\mathrm{f}$.
To compute the scattering from surfaces terminated at $\alpha$ and $\beta$ terraces, we terminate the bulk at a $\beta$ terrace, and incorporate an extra half unit cell of atoms in their bulk positions into the bottom of the reconstructed overlayer for the $\alpha$ terrace regions.
We also reverse the relaxation amounts in the $y$ direction for the $\alpha$ terraces, relative to those for the $\beta$ terraces.
Figure~\ref{fig:exact} illustrates these arrangements.
Appendix A and Supplemental Tables I-IX \cite{2021_Ju_PRB_supplemental} give values of atomic coordinates $\mathbf{r}_{jkn}^x$ used for the $x = \alpha$ and $\beta$ structure factors for each of the reconstructions considered.
For all calculations presented in this paper, we use a photon energy of $25.78$~keV ($\lambda = 0.4809$~\AA) to correspond with recent experiments \cite{2020_Ju_NC}. 
We use atomic form factors for each type of atom \cite{1995_Waasmaier_ActaCrystA51_416} with resonant corrections for this energy \cite{1993_Henke_ANDT54_181}, an absorption length of $\ell_\mathrm{abs} = 103$ $\mu$m for GaN at this energy, 
a Debye-Waller length of $u_k = 0.16$~\AA~ for all atoms,
and a surface roughness of $\sigma_R = 1$~\AA.

Figure~\ref{fig:CTR_exact} shows the calculated reflectivity as a function of $L$ for different integer $H_0 K_0$ values and three $\alpha$ fractions, $f_\alpha = 0$, $0.5$ and $1$.
Here we show calculations for an unreconstructed surface, with all atoms in their bulk positions.
The $(0 0 L)$ CTR is insensitive to the difference between the $\alpha$ and $\beta$ terminations ($f_\alpha = 1$ and $f_\alpha = 0$, respectively); 
both give the same intensity distribution.
However, it is very sensitive to whether the surface has a single termination ($\alpha$ or $\beta$), or a mixture of the two.
Deep minima form near the forbidden Bragg positions (odd $L_0$) for $f_\alpha = 0.5$ owing to destructive interference between the scattering from the $\alpha$ and $\beta$ regions.
We see sharp features exactly at forbidden Bragg peaks such as $L = 1$ on the $(0 0 L)$ CTR for all values of $f_\alpha$.
These arise from the effect of the absorption factor in $Z$ with a non-zero $\epsilon$.
In contrast to $(0 0 L)$, the $(0 2 L)$ and $(1 1 L)$ CTRs show very different intensity distributions for $\alpha$ and $\beta$ terminations.
There are alternating deep and shallow minima between the Bragg peaks, with the alternation being opposite for the two terminations.
Furthermore, the $(0 2 L)$ scattering from the $\alpha$ terrace is identical to the $(1 1 L)$ scattering from the $\beta$ terrace, and vice versa, as required by symmetry.
For surfaces with mixed terrace coverages, the $(0 2 L)$ CTR for $f_\alpha = X$ and the $(1 1 L)$ CTR for $f_\alpha = 1-X$ are similar but not identical.
This can be seen for the $f_\alpha = 0.5$ case shown, where the minima between the Bragg peaks all have about the same depth.
The small difference between the two CTRs is determined by whether the $\alpha$ terraces or $\beta$ terraces are the top layer.
The calculations shown here are for $\alpha$ islands on top.
By symmetry, the $(0 2 L)$ CTR for $f_\alpha = X$ with $\alpha$ islands is identical to the $(1 1 L)$ CTR for $f_\alpha = 1-X$ with $\beta$ islands.

We have performed calculations using atomic coordinates for all of the stable GaN $(0 0 1)$ reconstructions predicted previously \cite{2012_Walkosz_PRB_85_033308}.
On this surface, there are 6 reconstruction domain orientations, related by 3-fold rotation about the $6_3$ axis, and/or reflection about a plane passing through the axis at $x = 0$.
Figure~\ref{fig:CTR_exact_recon} shows calculations for $f_\alpha = 0$ with equal fractions $\theta_{xj} = 1/6$ of all six domains for each reconstruction. The label ``bulk'' refers to no reconstruction with all atoms in their bulk atomic positions; ``bare'' has relaxed atomic positions in the surface layers with no extra atoms (see Supplemental Table I \cite{2021_Ju_PRB_supplemental}); other reconstructions have relaxed positions and extra atoms at locations indicated (see Appendix A Table \ref{tab:coord_reconb} and \ref{tab:coord_recona} for the atomic coordinates of the 3H(T1) reconstruction and Supplemental Tables II to IX \cite{2021_Ju_PRB_supplemental} for the other 8 reconstructions).
All show the same qualitative behavior as the unreconstructed surface, with small quantitative differences.
Furthermore, because the X-ray scattering is dominated by the Ga atoms, which occupy an HCP lattice, the same qualitative behavior found for GaN is also obtained for an elemental HCP crystal.

\subsection{Miscut surface with alternating terminations}

\begin{figure}
\includegraphics[width=1\linewidth]{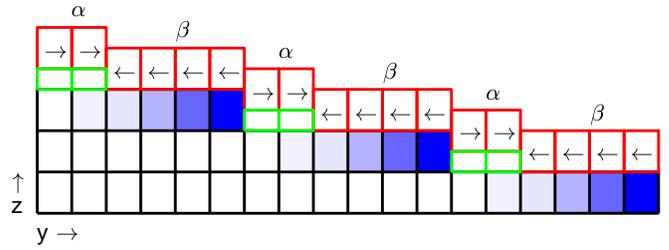}
\caption{Bulk unit cells (black), extra half unit cells (green) producing shift between $\alpha$ and $\beta$ terminations of neighboring terraces, and reconstructed unit cells (red), for a miscut surface with $M = 6$, $N = 2$, and $f_\alpha = 1/3$. Blue shade indicates index $m$ of sum in Eq.~(\ref{eq:r_b_v}), with final term $m = 0$ darkest. \label{fig:recon}}
\end{figure}

We now consider a miscut surface, with an array of straight steps parallel to the $x$ axis, periodically distributed along the $y$ axis, as shown in Fig.~\ref{fig:termination}.
We assume that the miscut corresponds to a decrease of the surface height by a full unit cell $c$ every $M$ unit cells in $y$, so that the period of the step array is $Mb$.
The surface miscut angle $\gamma$ relative to $(001)$ is given by $\tan{\gamma} = c / (Mb)$, and the surface is parallel to $(0 1 M)$ planes.
As shown in Fig.~\ref{fig:KL}(b), the CTRs from this surface are tilted in the $Q_y$ direction at an angle $\gamma$ from $(0 0 1)$.
Because of the tilt, there are $M$ times as many CTRs as in the exactly oriented case, indexed not just by $H_0 K_0$ but also by values of $L_0$ from $0$ to $M-1$.
The $Q_y$ value varies with $L$ along the CTR according to $Q_y = (2 \pi / b) [K_0 + (L-L_0)/M]$, where $H_0 K_0 L_0$ is the primary Bragg peak associated with the CTR.
The spacing in $L$ along a given CTR between Bragg peak positions is $M$, rather than unity as in the exactly oriented surface.
(This neglects systematic absences in the Bragg peaks; for the orthohexagonal case, the spacing between allowed peaks is $2M$.)

We assume that the surface has alternating terminations of $\alpha$ and $\beta$ terraces along the $y$ axis.
Figure \ref{fig:recon} shows the bulk and reconstructed unit cells used to calculate the CTRs for the vicinal surface.
The width of the $\alpha$ terraces is $N$ unit cells, and the width of the $\beta$ terraces is $M - N$ unit cells.
The $\alpha$ terrace fraction is given by $f_\alpha = N / M$.

To calculate the scattering from a miscut crystal, it is useful to take advantage of the new periodicity in the surface plane produced by the step array \cite{2002_Trainor_JApplCryst35_696}.
The bulk reflectivity amplitude can be built up by summing the scattering from each unit cell in the $y$ direction from $m = -\infty$ to $m = 0$, according to
\begin{equation}
    r_\mathrm{bulk} = \frac{r_\mathrm{f} \, F_\mathrm{bulk}}{M} \sum_{m = -\infty}^0 Y^m = \frac{r_\mathrm{f} \, F_\mathrm{bulk}}{M} \frac{Y}{Y - 1}, \label{eq:r_b_v} 
\end{equation}
where the quantity $Y$ is the ratio of the scattering amplitude from one unit cell to that from the unit cell at $\Delta y = -b$ beside it, made up of a phase factor and an absorption factor
\begin{equation}
    Y \equiv \exp (i Q_y b) \, \exp(\epsilon \, b \tan \gamma / Q_z).
\end{equation}
This is similar to the geometric sequence for the exactly oriented surface, Eq.~(\ref{eq:r_b}), except that the summation is carried out in the $y$ direction rather than the $z$ direction, as indicated in Fig.~\ref{fig:recon}.
Substituting the above expression for $Q_y$ along the CTR into the phase factor, the term involving $K_0$ drops out.
This gives
\begin{align}
    Y &= \exp [i (Q_z c - 2 \pi L_0)/M] \exp (\epsilon c/M Q_z ) \nonumber \\
    &= \exp [2 \pi i (L - L_0) / M] \exp (\epsilon c^2 / 2 \pi M L).
\end{align}
Comparison with Eq.~(\ref{eq:Z}) shows that $Y^M = Z$.

The reflectivity amplitude per unit area of the reconstructed overlayer on the $\alpha$ terraces can be written as
\begin{equation}
    r_\mathrm{rec}^\alpha = \frac{r_\mathrm{f} \, F_\mathrm{rec}^\alpha}{N} \sum_{m = 1}^{N} Y^m = \frac{r_\mathrm{f} \, F_\mathrm{rec}^\alpha}{N} \frac{Y(Y^N - 1)}{Y - 1}.
    \label{eq:ra}
\end{equation}
A similar expression applies to the $\beta$ terraces,
\begin{equation}
    r_\mathrm{rec}^\beta = \frac{r_\mathrm{f} \, F_\mathrm{rec}^\beta}{M-N} \sum_{m = N + 1}^{M} Y^m = \frac{r_\mathrm{f} \, F_\mathrm{rec}^\beta}{M-N} \frac{Y(Y^M - Y^N)}{Y - 1}.
     \label{eq:rb}
\end{equation}
In these equations the sums run over the fraction of the surface periodicity covered by each termination.

As in the calculation for the exactly oriented surface, the total reflectivity amplitude $r_\mathrm{tot}$ is the sum of the complex amplitudes from the bulk and the reconstructed layers on the $\alpha$ and $\beta$ terraces.
Expressions similar to Eqs.~(\ref{eq:rt}-\ref{eq:R}) give the intensity reflectivity $R_{L_0}$ for the CTR from the $H_0 K_0 L_0$ Bragg peak, with the roughness factor for each CTR on a vicinal surface now given by a simple Gaussian form \cite{1985_Andrews_JPhysC18_6427,1997_Munkholm_JApplPhys82_2944,2017_Petach_PRB95_184104},
\begin{equation}
    S_{L_0} = \exp \big [ - \sigma_R^2 (2\pi/c)^2 (L - L_0)^2 \big ],
    \label{eq:SL0}
\end{equation}
where we assume small miscut ($M$ large) so that only the Bragg peak at $L = L_0$ need be considered for each CTR.
Away from Bragg peaks, so that the dynamical scattering correction can be neglected, the intensity reflectivity can be written as
\begin{align}
    R_{L_0} &= \left | \frac{r_\mathrm{f}}{M} \right |^2 \, |Y|^2 \,\, S_{L_0} \label{eq:RL0} \\
    &\times \left | \frac{F_\mathrm{bulk} + F_\mathrm{rec}^\alpha (Y^N - 1) + F_\mathrm{rec}^\beta (Y^M - Y^N)}{Y - 1} \right |^2. \nonumber
\end{align}

\subsection{Miscut GaN $(0 0 1)$ surface}

\begin{figure*}
\centering
\includegraphics[width=0.75\linewidth]{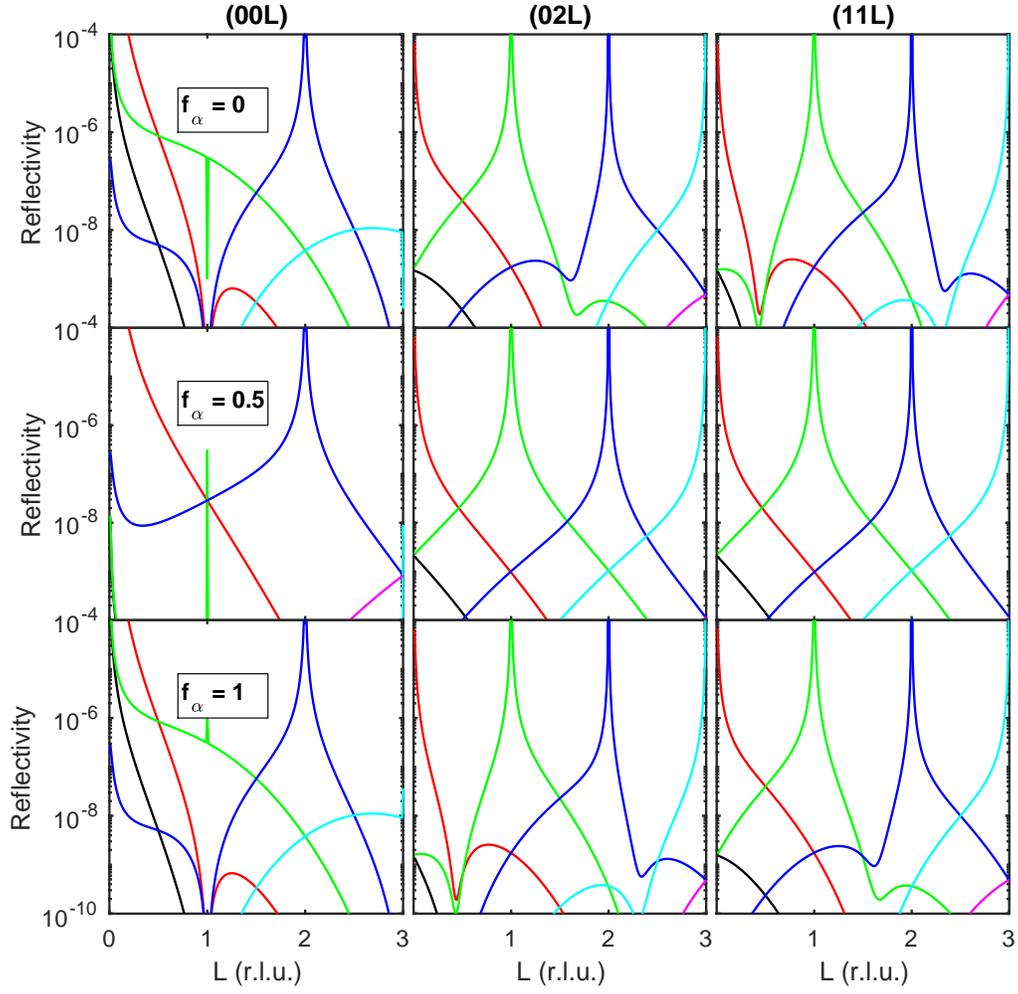} 
\caption{Calculated reflectivities of CTRs for a miscut GaN $(0 0 1)$ surface with $M = 100$ and no reconstruction. Left column: $(0 0 L)$; middle column: $(0 2 L)$; right column: $(1 1 L)$. Black, red, green, blue, cyan, and magenta curves are for $L_0 = -1$ to $4$, respectively. Values of $f_\alpha$ are given for each row. \label{fig:CTR_miscut} }
\end{figure*}

\begin{figure*}
\includegraphics[width=1.0\linewidth]{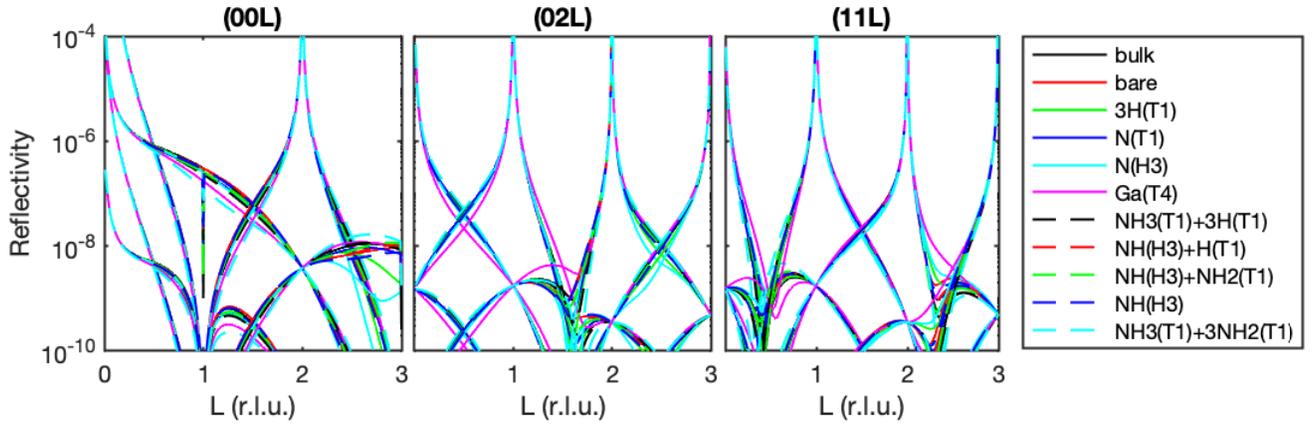}
\caption{Calculated reflectivity of CTRs for a miscut GaN $(0 0 1)$ surface with $M = 100$ and various reconstructions, for $f_\alpha = 0$. . \label{fig:CTR_miscut_recon}}
\end{figure*}

\begin{figure*}
\includegraphics[width=0.9\linewidth]{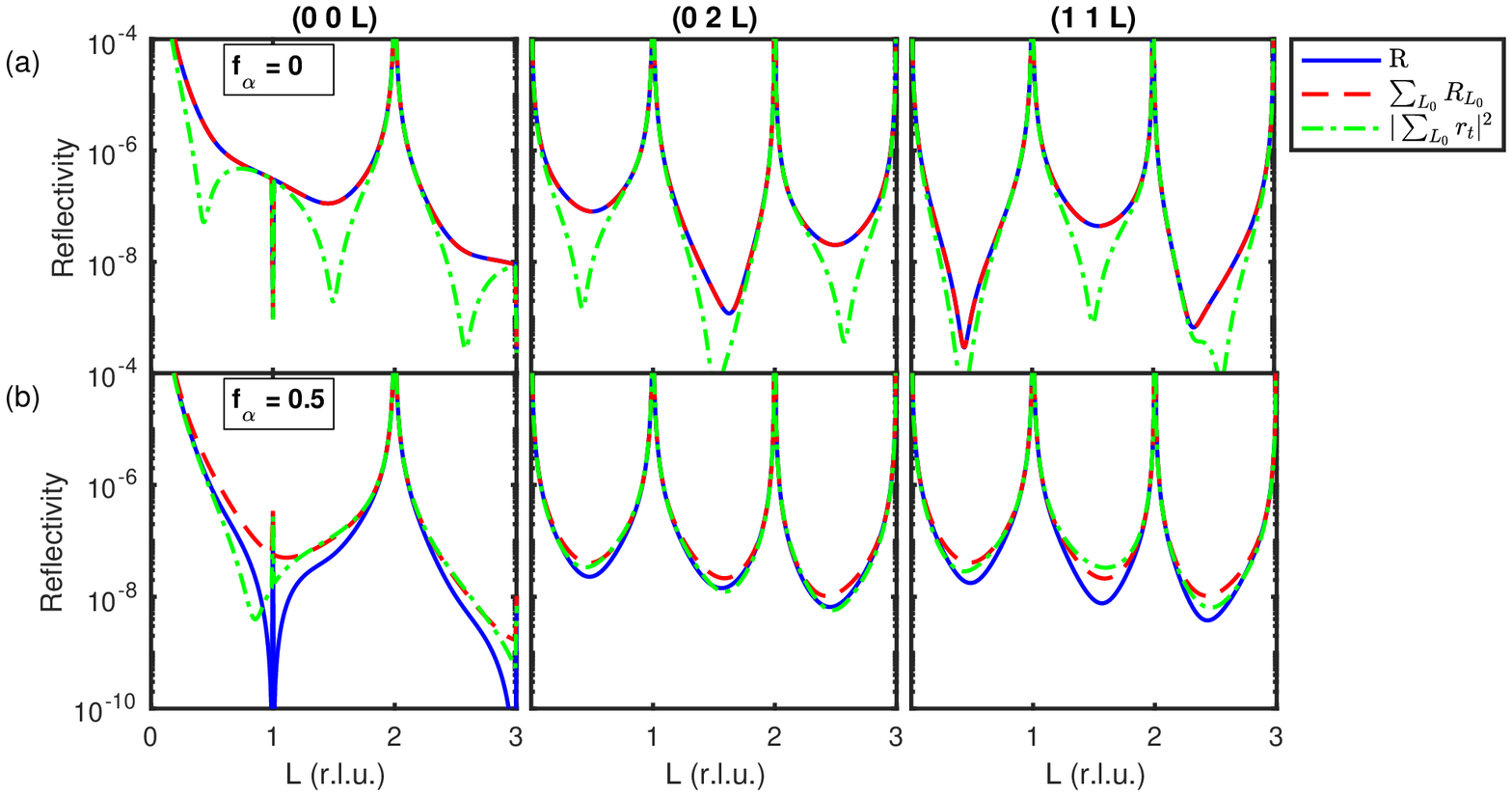}
\caption{Relationship of CTR profiles for an exactly oriented surface (solid blue curves) to those for miscut surfaces, for GaN $(0 0 1)$ with no reconstruction. Red and green dashed curves show sum over $L_0$ of intensities and square of sum of amplitudes, respectively, of all CTRs at the same $H_0 K_0$ from a miscut surface. (a) and (b) correspond to single and mixed terminations, $f_\alpha = 0$ and $0.5$, respectively. \label{fig:CTR_both_0}}
\end{figure*}

Figure \ref{fig:CTR_miscut} shows the calculated reflectivity of various CTRs as a function of $L$ for a miscut GaN $(0 0 1)$ surface. 
Three families of CTRs are shown, $(0 0 L_0)$, $(0 2 L_0)$, and $(1 1 L_0)$, for $L_0 = -1$ to $4$, with three $f_\alpha$ values, 0.0, 0.5, and 1.0.
These calculations were done for a step period of $M = 100$ and an unreconstructed surface.
The result is insensitive to changes in $M$; any value above $M = 10$ gives reflectivities that agree within $\sim 10 \%$ at all $L$.
As in the case of an exactly oriented surface, the $(0 0 L_0)$ CTRs are identical for $f_\alpha = 0$ and $f_\alpha = 1$, and they are very different for $f_\alpha = 0.5$, with the CTRs for even $L_0$ becoming stronger and the CTRs for odd $L_0$ becoming very weak.
The $(0 2 L_0)$ and $(1 1 L_0)$ CTRs have a more monotonic dependence on $f_\alpha$.
For $f_\alpha = 0$ and $f_\alpha = 1$, there are alternating stronger and weaker intensities between the Bragg peaks, with the alternation being opposite for $(0 2 L_0)$ and $(1 1 L_0)$.
For $f_\alpha = 0.5$, the intensities between the Bragg peaks are about the same, and there is no difference between the $(0 2 L_0)$ and $(1 1 L_0)$ CTRs.
The $(0 2 L_0)$ CTRs with $f_\alpha = X$ are identical to the $(1 1 L_0)$ CTRs with $f_\alpha = 1-X$, for any value $X$.
As with the exactly oriented surface, other GaN reconstructions have the same qualitative behavior, as shown in Fig.~\ref{fig:CTR_miscut_recon}, as do elemental HCP structures. The calculations for Fig.~\ref{fig:CTR_miscut_recon} use the same atomic coordinates for the reconstructions as in Fig.~\ref{fig:CTR_exact_recon}.

\subsection{Relation between CTR profiles for miscut and exactly oriented surfaces}

It has been previously discussed that under some circumstances the CTR intensity profile for an exactly oriented surface can be expressed as the sum of the CTR intensities from each Bragg peak along the rod \cite{1990_FaradayDiscussion,1997_Munkholm_JApplPhys82_2944}.
This is analogous to the sum of the intensities of all the CTRs in the family having the same $H_0 K_0$ from a miscut surface.
Here we make this comparison with the expressions obtained above for the cases of surfaces with pure $\alpha$, pure $\beta$, or mixed terminations.

For a miscut surface with a single termination, i.e. $f_\alpha = 0$ or $1$, giving full-unit-cell-height steps, one can show that the intensity reflectivity $R$ for an exactly oriented surface is simply the sum of the intensity reflectivities $R_{L_0}$ for all the CTRs with the same $H_0 K_0$, in the limit of small miscut angle and neglecting absorption,
\begin{equation}
    R = \lim_{M \rightarrow \infty} \sum_{L_0} R_{L_0}.
    \label{eq:sum}
\end{equation}
For $N = 0$ or $N = M$ ($x = \beta$ or $\alpha$ termination, respectively), using $Y^M = Z$, Eq.~(\ref{eq:RL0}) becomes
\begin{equation}
    R_{L_0} = \left | \frac{r_\mathrm{f}}{M} \right |^2 \, |Y|^2 \, \left | \frac{F_\mathrm{bulk}}{Z - 1} + F_\mathrm{rec}^x \right |^2 \, \left | \frac{Z - 1}{Y - 1} \right |^2 S_{L_0}.
    \label{eq:RL02}
\end{equation}
Neglecting the absorption factors in $Z$ and $Y$, one obtains $|Z - 1|^2 = 4 \sin^2(\pi L)$ and $|Y - 1|^2 = 4 \sin^2[\pi (L - L_0)/M]$. 
For small miscut angle (large $M$), the second expression becomes $|Y - 1|^2 = 4 \pi^2 (L - L_0)^2/M^2$.
Substituting these into Eq.~(\ref{eq:RL02}) gives
\begin{equation}
    \lim_{M \rightarrow \infty} R_{L_0} = |r_\mathrm{f}|^2 \, |Y|^2 \, \left | \frac{F_\mathrm{bulk}}{Z - 1} + F_\mathrm{rec}^x \right |^2 \frac{\sin^2(\pi L)}{\pi^2 (L - L_0)^2} S_{L_0}.
    \label{eq:RL03}
\end{equation}
While the first two factors depend upon $Q$ and thus vary slightly with $L_0$ at fixed $L$ due to the spacing of the CTRs in $K$, for small miscut angle this is negligible, and these factors are the same as for an exactly oriented surface.
Comparison with Eqs.~(\ref{eq:R}), (\ref{eq:S}), and (\ref{eq:SL0}) shows that the summation Eq.~(\ref{eq:sum}) holds, if we neglect the factors of $|Z|^2$ and $|Y|^2$, which can be set to unity when absorption is small.

This is illustrated in Fig.~\ref{fig:CTR_both_0}(a), where the summation over $L_0$ of the miscut CTR intensities $R_{L_0}$ gives the same result as the calculation of $R$ for the exactly oriented surface for $f_\alpha = 0$.
This equality holds for all CTRs, even for those such as $(0 2 L)$ and $(1 1 L)$ which are sensitive to the differences between the $\alpha$ and $\beta$ terminations.
This result differs from previous analyses \cite{1990_FaradayDiscussion,1997_Munkholm_JApplPhys82_2944} because we include the $L$ dependence of the structure factors $F_\mathrm{bulk}$ and $F_\mathrm{rec}$, which properly accounts for the sensitivity of the CTRs to the surface termination in the regions between the Bragg peaks.
However, as shown in Fig.~\ref{fig:CTR_both_0}(b), the summation Eq.~(\ref{eq:sum}) does \textit{not} hold for mixed terminations, e.g. $f_\alpha = 0.5$. 

It is interesting that, for the single termination surface, the exactly oriented CTR is equal to the sum of the \textit{intensities} of the miscut CTRs, not the square of the sum of the \textit{amplitudes} of the miscut CTRs.
This is demonstrated in Fig.~\ref{fig:CTR_both_0}(a).
From an experimental perspective, this means that there is no difference between the CTR profile measured from an exactly oriented surface and that from a surface with a miscut so small that the splitting of the CTRs cannot be resolved.
For the mixed termination surface, Fig.~\ref{fig:CTR_both_0}(b), the exactly oriented CTR is not equal to the miscut CTRs summed in either intensity or amplitude.
In this case the reflectivity of the exactly oriented surface depends upon whether the $\alpha$ or $\beta$ islands are on top, and the limit of zero miscut $M \rightarrow \infty$ does not reduce the miscut surface structure to that of the exactly oriented surface.

\section{Discussion}

The expressions developed above are an extension of those developed previously \cite{2017_Petach_PRB95_184104} to the case of alternating terrace structures, while also including finite absorption and orthorhombic symmetry.
The kinematic expression for the CTR intensity as a function of $L$ for a miscut surface, Eq.~(\ref{eq:RL0}), is equivalent to the second line of Eq.~(12) in \cite{2017_Petach_PRB95_184104} if we neglect absorption so that $Y = \exp[2 \pi i (L - L_0)/M]$ and $|Y - 1|^2 = 4 \sin^2[\pi (L - L_0)/M]$, consider a single termination ($N = 0$ or $N = M$), and cubic symmetry $c = b = a$.
Likewise the kinematic expression for an exactly oriented surface, Eq.~(\ref{eq:R2}), is equivalent to Eq.~(17) in \cite{2017_Petach_PRB95_184104} under the same conditions.

The analysis presented above shows that CTR profiles are very sensitive to the difference between $\alpha$ and $\beta$ terraces on vicinal basal plane surfaces of HCP-type systems.
This enables \textit{in situ} X-ray measurements during growth to determine the surface fraction covered by each type of terrace, and thus to distinguish the dynamics of $A$ and $B$ steps.
Measurement of the CTR profiles as a function of $L$ can be carried out by running scans during steady-state growth conditions.
Recent measurements during step-flow growth by organo-metallic vapor phase epitaxy of GaN $(0 0 0 1)$ \cite{2020_Ju_NC} show CTR profiles that agree with the calculations presented here, allowing the steady-state terrace fraction $f_\alpha$ to be determined as a function of growth conditions.
The observed increase in $f_\alpha$ with increasing growth rate indicates that adatom attachment kinetics are faster at $A$ steps than $B$ steps for this system.

\begin{figure}
\includegraphics[width=0.9\linewidth]{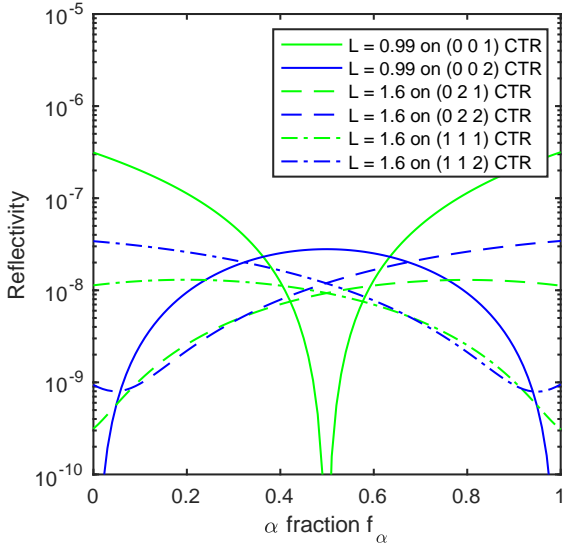}
\caption{Calculated reflectivity of selected CTRs for a miscut GaN $(0 0 1)$ surface as a function of terrace fraction $f_\alpha$, for fixed values of $L$ given.  \label{fig:R_vs_f}}
\end{figure}

\begin{figure}
\includegraphics[width=0.9\linewidth]{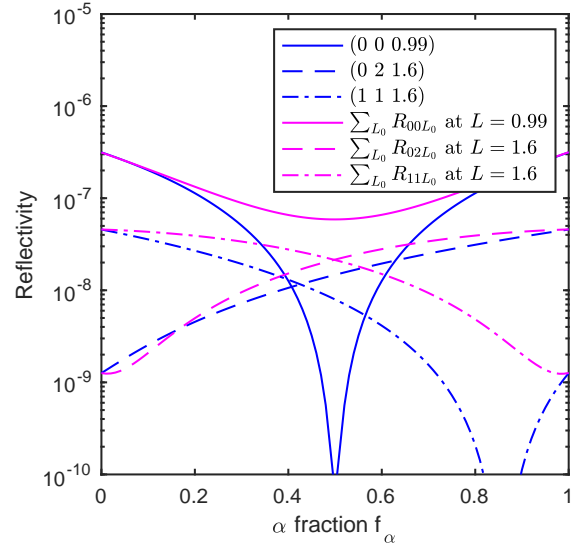}
\caption{Blue curves show calculated reflectivity of selected CTRs for an exactly oriented GaN $(0 0 1)$ surface as a function of terrace fraction $f_\alpha$, for fixed values of $L$ given. Magenta curves show sum over $L_0$ of intensities of all CTRs at the same $H_0 K_0$ from a miscut surface. \label{fig:Rz_vs_f}}
\end{figure}

In addition, the dynamics of the terrace fraction $f_\alpha(t)$ can be observed by monitoring CTR intensities at particular $L$ values during changes in growth conditions.
Figure \ref{fig:R_vs_f} shows calculations of the reflectivity as a function of $f_\alpha$ at fixed $L$ positions on six CTRs, for a miscut unreconstructed GaN $(0 0 1)$ surface.
On the $(0 0 1)$ CTR at $L = 0.99$, the intensity is maximum for $f_\alpha = 0$ or $f_\alpha = 1$, and goes through a deep minimum at $f_\alpha = 0.5$.
On the $(0 0 2)$ CTR at $L = 0.99$, the intensity is minimum for $f_\alpha = 0$ or $f_\alpha = 1$, and goes through a maximum at $f_\alpha = 0.5$.
On the $(0 2 1)$ and $(0 2 2)$ CTRs at $L = 1.6$, the intensity increases almost monotonically as $f_\alpha$ increases from $0$ to $1$.
The $(1 1 1)$ and $(1 1 2)$ CTRs at $L = 1.6$ show complementary behavior, with an almost monotonic intensity decrease.
These curves can be used to convert observed intensity changes into terrace fraction dynamics $f_\alpha(t)$, e.g. as net growth rate is changed during step-flow growth \cite{2020_Ju_NC}.

Figure \ref{fig:Rz_vs_f} shows calculations of the reflectivity as a function of $f_\alpha$ at fixed $L$ positions on three CTRs for an exactly oriented unreconstructed GaN $(0 0 1)$ surface.
The $(0 0 L)$ with $L = 0.99$ position shows behavior qualitatively similar to the $(0 0 1)$ CTR of a miscut surface, with a minimum at $f_\alpha = 0.5$.
This position (or the equivalent $(1 3 L)$ position) has been used to monitor growth under island nucleation and coalescence conditions, where oscillations in intensity occur with a period corresponding to growth of half-unit-cell monolayers \cite{1999_Stephenson_APL74_3326,2014_Perret_APL105_051602}.
Figure \ref{fig:Rz_vs_f} also shows how the sum of the CTR intensities for a miscut surface, which agree with the exactly oriented surface at $f_\alpha = 0$ and $1$, deviate at intermediate values of $f_\alpha$.

\section{Conclusions and Outlook}

The above analysis extends previous calculations of CTR profiles for miscut surfaces to include half-unit-cell-height steps and variation of the coverage of the $\alpha$ and $\beta$ terraces above each step.
We demonstrate that the CTR intensity as a function of $L$ can be used to measure the terrace fraction $f_\alpha$, and thus determine whether $A$ or $B$ steps have faster kinetics during crystal growth on basal plane surfaces of HCP-type systems.
In particular the orthohexagonal $(0 2 L)$ and $(1 1 L)$ CTRs (equivalent to $(0 1 \overline{1} L)$ and $(1 0 \overline{1} L)$, respectively, in Miller-Bravais notation) show characteristic alternating high and low intensities between the Bragg peaks that depend on $f_\alpha$.
Intensities between Bragg peaks e.g. at $L = 1.6$ on the $(0 2 2)$ or $(1 1 2)$ CTRs change almost monotonically with $f_\alpha$.
This behavior is qualitatively similar for different GaN $(0 0 1)$ surface reconstructions, and for elemental HCP systems.
A comparison of CTR profiles for exactly oriented and miscut surfaces shows that a summation relation Eq.~(\ref{eq:sum}) holds for all CTRs, but only when the surface has a single termination, $f_\alpha = 0$ or $1$. 

While the example CTR calculations presented here are for wurtzite-structure GaN, the ability to distinguish the terrace fraction $f_\alpha$ applies directly to many other HCP-type systems with a $6_3$ screw axis, including other compound semiconductors, as well as one third of the crystalline elements and many more complex crystals.
The treatment for miscut surfaces developed above considers only two terminations that alternate, separated by half-unit-cell-height steps, to provide a method to distinguish $A$ and $B$ step properties in HCP-type systems.
For crystal structures with other types of screw axes or multiple layers \cite{2004_Fenter_JApplCryst37_977}, the geometrical sequence formulas Eqs.~(\ref{eq:ra}-\ref{eq:rb}) are straightforward to extend to periodic sequences of more than two terminations.

\begin{acknowledgments}
  Work supported by the U.S Department of Energy (DOE), Office of Science, Office of Basic Energy Sciences, Materials Science and Engineering Division.
  This work was inspired by and developed to analyze data from experiments at beamline 12ID-D of the Advanced Photon Source, a DOE Office of Science user facility operated by Argonne National Laboratory.
\end{acknowledgments}

\appendix
\section{Atomic coordinates}

\begin{table}
\caption{ \label{tab:coord_sub} Fractional orthohexagonal coordinates of bulk GaN used to calculate the substrate contribution to the CTRs.}
\begin{ruledtabular}
\begin{tabular}{ c | c || r | r | r }
Atom & Site & $x$ & $y$ & $z$\\
$k$ & $n$ & & & \\
\hline
Ga & 1 & 0.5000  &  0.1667 &  -0.5000 \\
Ga & 2 & 0.0000  &  0.6667 &  -0.5000 \\
Ga & 3 & 1.5000  &  0.1667 &  -0.5000 \\
Ga & 4 & 1.0000  &  0.6667 &  -0.5000 \\
Ga & 5 & 0.0000  &  0.0000 &   0.0000 \\
Ga & 6 & 0.5000  &  0.5000 &   0.0000 \\
Ga & 7 & 1.0000  &  0.0000 &   0.0000 \\
Ga & 8 & 1.5000  &  0.5000 &   0.0000 \\
\hline
N & 1 & 0.0000   &  0.0000  &  -0.6232 \\
N & 2 & 0.5000   &  0.5000  &  -0.6232 \\
N & 3 & 1.0000   &  0.0000  &  -0.6232 \\
N & 4 & 1.5000   &  0.5000  &  -0.6232 \\
N & 5 & 0.5000   &  0.1667  &  -0.1232 \\
N & 6 & 0.0000   &  0.6667  &  -0.1232 \\
N & 7 & 1.5000   &  0.1667  &  -0.1232 \\
N & 8 & 1.0000   &  0.6667  &  -0.1232 \\
\end{tabular}
\end{ruledtabular}
\end{table}

\begin{table} 
\caption{\label{tab:coord_recona} Fractional orthohexagonal coordinates $x$, $y$, $z$ of atoms in domain $j = 1$ of the 3H(T1) reconstruction used to calculate the $\alpha$ terrace contribution to the CTRs, as well as their differences $\Delta x$, $\Delta y$, $\Delta z$ relative to bulk lattice positions. The differences for H atoms are relative to N sites. The lowest four Ga and N sites are an extra half unit cell of bulk lattice to account for the difference in height of the $\alpha$ and $\beta$ terraces.}
\begin{ruledtabular}
\begin{tabular}{ c | c || r | r | r | r | r | r }
Atom & Site & $x$ & $y$ & $z$ & $\Delta x$ & $\Delta y$ & $\Delta z$\\
$k$ & $n$ & & & & & &\\
\hline
Ga & 1 & 0.5000  &   0.1667  &  -0.5000  &   0.0000  &   0.0000  &   0.0000 \\
Ga & 2 & 0.0000  &   0.6667  &  -0.5000  &   0.0000  &   0.0000  &   0.0000 \\
Ga & 3 & 1.5000  &   0.1667  &  -0.5000  &   0.0000  &   0.0000  &   0.0000 \\
Ga & 4 & 1.0000  &   0.6667  &  -0.5000  &   0.0000  &   0.0000  &   0.0000 \\
Ga & 5 & 0.0000  &   0.0000  &   0.0076  &   0.0000  &   0.0000  &   0.0076 \\
Ga & 6 & 0.5075  &   0.4975  &  -0.0015  &   0.0075  &  -0.0025  &  -0.0015 \\
Ga & 7 & 1.0000  &   0.0050  &  -0.0015  &   0.0000  &   0.0050  &  -0.0015 \\
Ga & 8 & 1.4925  &   0.4975  &  -0.0015  &  -0.0075  &  -0.0025  &  -0.0015 \\
Ga & 9 & 0.4929  &   0.1643  &   0.5223  &  -0.0071  &  -0.0024  &   0.0223 \\
Ga &10 & 0.0000  &   0.6667  &   0.4294  &   0.0000  &   0.0000  &  -0.0706 \\
Ga &11 & 1.5071  &   0.1643  &   0.5223  &   0.0071  &  -0.0024  &   0.0223 \\
Ga &12 & 1.0000  &   0.6714  &   0.5223  &   0.0000  &   0.0047  &   0.0223 \\
\hline
N & 1 & 0.0000  &   0.0000  &  -0.6232  &   0.0000  &   0.0000  &   0.0000 \\
N & 2 & 0.5000  &   0.5000  &  -0.6232  &   0.0000  &   0.0000  &   0.0000 \\
N & 3 & 1.0000  &   0.0000  &  -0.6232  &   0.0000  &   0.0000  &   0.0000 \\
N & 4 & 1.5000  &   0.5000  &  -0.6232  &   0.0000  &   0.0000  &   0.0000 \\
N & 5 & 0.4988  &   0.1663  &  -0.1254  &  -0.0012  &  -0.0004  &  -0.0022 \\
N & 6 & 0.0000  &   0.6667  &  -0.1201  &   0.0000  &   0.0000  &   0.0031 \\
N & 7 & 1.5012  &   0.1663  &  -0.1254  &   0.0012  &  -0.0004  &  -0.0022 \\
N & 8 & 1.0000  &   0.6675  &  -0.1254  &   0.0000  &   0.0008  &  -0.0022 \\
N & 9 & 0.0000  &   0.0000  &   0.3766  &   0.0000  &   0.0000  &  -0.0002 \\
N &10 & 0.5064  &   0.4979  &   0.3775  &   0.0064  &  -0.0021  &   0.0007 \\
N &11 & 1.0000  &   0.0043  &   0.3775  &   0.0000  &   0.0043  &   0.0007 \\
N &12 & 1.4936  &   0.4979  &   0.3775  &  -0.0064  &  -0.0021  &   0.0007 \\
\hline
H &13 & 0.5002  &   0.1667  &   0.8196  &   0.0002  &   0.0000  &  -0.0572 \\
H &15 & 1.4998  &   0.1667  &   0.8196  &  -0.0002  &   0.0001  &  -0.0572 \\
H &16 & 1.0000  &   0.6666  &   0.8196  &   0.0000  &  -0.0001  &  -0.0572 \\

\end{tabular}
\end{ruledtabular}
\end{table}

\begin{table} 
\caption{\label{tab:coord_reconb} Fractional orthohexagonal coordinates $x$, $y$, $z$ of atoms in domain $j = 1$ of the 3H(T1) reconstruction used to calculate the $\beta$ terrace contribution to the CTRs, as well as their differences $\Delta x$, $\Delta y$, $\Delta z$ relative to bulk lattice positions. The differences for H atoms are relative to N sites.}
\begin{ruledtabular}
\begin{tabular}{ c | c || r | r | r | r | r | r }
Atom & Site & $x$ & $y$ & $z$ & $\Delta x$ & $\Delta y$ & $\Delta z$\\
$k$ & $n$ & & & & & &\\
\hline
Ga & 1 & 0.5075  &   0.1692  &  -0.5015  &   0.0075  &   0.0025  &  -0.0015 \\
Ga & 2 & 0.0000  &   0.6667  &  -0.4924  &   0.0000  &   0.0000  &   0.0076 \\
Ga & 3 & 1.4925  &   0.1692  &  -0.5015  &  -0.0075  &   0.0025  &  -0.0015 \\
Ga & 4 & 1.0000  &   0.6617  &  -0.5015  &   0.0000  &  -0.0050  &  -0.0015 \\
Ga & 5 & 0.0000  &   0.0000  &  -0.0706  &   0.0000  &   0.0000  &  -0.0706 \\
Ga & 6 & 0.4929  &   0.5024  &   0.0223  &  -0.0071  &   0.0024  &   0.0223 \\
Ga & 7 & 1.0000  &  -0.0047  &   0.0223  &   0.0000  &  -0.0047  &   0.0223 \\
Ga & 8 & 1.5071  &   0.5024  &   0.0223  &   0.0071  &   0.0024  &   0.0223 \\
\hline
N & 1 & 0.0000  &   0.0000  &  -0.6201  &   0.0000  &   0.0000  &   0.0031 \\
N & 2 & 0.4988  &   0.5004  &  -0.6254  &  -0.0012  &   0.0004  &  -0.0022 \\
N & 3 & 1.0000  &  -0.0008  &  -0.6254  &   0.0000  &  -0.0008  &  -0.0022 \\
N & 4 & 1.5012  &   0.5004  &  -0.6254  &   0.0012  &   0.0004  &  -0.0022 \\
N & 5 & 0.5064  &   0.1688  &  -0.1225  &   0.0064  &   0.0021  &   0.0007 \\
N & 6 & 0.0000  &   0.6667  &  -0.1234  &   0.0000  &   0.0000  &  -0.0002 \\
N & 7 & 1.4936  &   0.1688  &  -0.1225  &  -0.0064  &   0.0021  &   0.0007 \\
N & 8 & 1.0000  &   0.6624  &  -0.1225  &   0.0000  &  -0.0043  &   0.0007 \\
\hline
H &10 & 0.5002  &   0.5000  &   0.3196  &   0.0002  &  -0.0000  &  -0.0572 \\
H &11 & 1.0000  &   0.0001  &   0.3196  &   0.0000  &   0.0001  &  -0.0572 \\
H &12 & 1.4998  &   0.4999  &   0.3196  &  -0.0002  &  -0.0001  &  -0.0572 \\
\end{tabular}
\end{ruledtabular}
\end{table}

To provide a detailed example of how we calculate the CTR intensities including the effects of reconstruction, we here provide an example of the calculated atomic coordinates we use for a particular GaN (0001) reconstruction, 3HT1.
Atomic coordinates calculated for the other reconstructions are given in Supplemental Tables I to IX \cite{2021_Ju_PRB_supplemental}.
All coordinates were obtained by \textit{ab initio} methods in \cite{2012_Walkosz_PRB_85_033308}.
The qualitative behavior we observe does not depend upon the reconstruction chosen or the exact values of the atomic coordinates used.

We first show the coordinates used for bulk positions in Table~\ref{tab:coord_sub}.
The fractional coordinates $x$, $y$, and $z$ given are the components of the positions $\mathbf{r}_{kn}^\mathrm{bulk}$ used to calculate $F_\mathrm{bulk}$, normalized to the respective orthohexagonal lattice parameters $a$, $b$, and $c$, i.e. $\mathbf{r} = (ax,by,cz)$.
A $2 \times 2$ surface unit cell is used, equivalent to two orthohexagonal unit cells, so there are 8 Ga and 8 N sites in each unit cell thickness.
These coordinates place a bulk Ga site on a $\beta$ layer at the origin;
$\beta$ terraces occur at integer values of $z$, while $\alpha$ terraces occur at half-integer values.
We use $u = 0.3768$ for the internal lattice parameter of bulk GaN, i.e. the fractional distance between Ga and N sites, which deviates slightly from the ideal $3/8$ value, in agreement with \textit{ab initio} calculations \cite{2012_Walkosz_PRB_85_033308,1999_Stampfl_PRB59_5521} and experiments \cite{2015_Minikayev_XraySpect44_382}.  

Tables~\ref{tab:coord_recona} and \ref{tab:coord_reconb} give the atomic coordinates for the 3H(T1) reconstruction. The fractional coordinates $x$, $y$, and $z$ given in the tables are the normalized components of the positions $\mathbf{r}_{jkn}^\alpha$ and $\mathbf{r}_{jkn}^\beta$ used to calculate $F_\mathrm{rec}^\alpha$ and $F_\mathrm{rec}^\beta$.  
Relaxed positions were calculated for a one-unit-cell thick layer at the surface.
For the $\alpha$ terrace, an extra half unit cell of bulk (unrelaxed) atoms is attached to the bottom to account for the difference in height of the $\alpha$ and $\beta$ terraces, as shown in Fig.~\ref{fig:recon}.
Coordinates for only one domain are given.
Those for other 5 domains are obtained by 3-fold rotation about the $6_3$ axis and/or reflection of the $y$ coordinate.
One can see that the Ga atoms bonded to the three adsorbed hydrogens of the 3H(T1) reconstruction relax to higher $z$ positions.

\end{document}